\documentclass[conference]{IEEEtran}
\IEEEoverridecommandlockouts
\usepackage{cite}
\usepackage{amsmath,amssymb,amsfonts}
\usepackage{algorithmic}
\usepackage{graphicx}
\usepackage{textcomp}
\usepackage{xcolor}
\usepackage{booktabs}
\usepackage{multirow}
\usepackage{url}
\def\BibTeX{{\rm B\kern-.05em{\sc i\kern-.025em b}\kern-.08em
    T\kern-.1667em\lower.7ex\hbox{E}\kern-.125emX}}
\begin{document}

\title{Automated Enrichment Analysis with GraphRAG: Efficient Multi-Database ORA Extraction and Biological Interpretation}

\author{
\IEEEauthorblockN{Guangcan Li\IEEEauthorrefmark{1},
Shuyuan Zheng\IEEEauthorrefmark{1},
Jiayao Liu\IEEEauthorrefmark{2},
Shigeto Seno\IEEEauthorrefmark{1},
Makoto Onizuka\IEEEauthorrefmark{1}}
\IEEEauthorblockA{\IEEEauthorrefmark{1}Graduate School of Information Science and Technology, The University of Osaka, Japan\\
\{guangcan.li, zheng, senoo, onizuka\}@ist.osaka-u.ac.jp}
\IEEEauthorblockA{\IEEEauthorrefmark{2}The First Affiliated Hospital of China Medical University, China\\
liujiayao@cmu.edu.cn}
}

\maketitle

\begin{abstract}
Over-representation analysis (ORA) is widely used to interpret gene-expression studies, but its outputs are often fragmented across gene-set databases and presented as flat term lists that obscure gene-level links among related enrichment signals. We propose Graph-Based Over-Representation Retrieval-Augmented Generation (GBOR-RAG), an integrated and traceable framework for ORA interpretation. GBOR-RAG represents heterogeneous gene-set databases as a unified term--gene knowledge graph, where shared genes connect related terms across sources. After graph-based ORA scoring, the framework retrieves a compact connected evidence subgraph that preserves both cross-database relationships and gene-level support. The subgraph conditions an LLM to generate biological interpretations, and we introduce a claim-level faithfulness metric to assess whether generated claims are supported by explicit evidence. Experiments show that graph-based ORA preserves conventional ORA results with high consistency (mean Pearson $r=0.997$) while reducing query time by a median of $37\times$ (up to $48\times$). The retrieved subgraphs average 36.8 nodes and achieve the highest overall integration score (4.12), and GBOR-RAG attains the highest claim-level faithfulness (0.90) among the compared systems. Adding a PubMed literature layer further improves the overall interpretation quality from 4.17 to 4.75.
\end{abstract}


\section{Introduction}
Over-representation analysis (ORA) is one of the most widely used methods for interpreting gene-expression profiles: it tests whether predefined gene sets are over-represented among differentially expressed genes relative to a background universe \cite{b2}. ORA relies on a rich ecosystem of curated gene-set resources, including Gene Ontology (GO) \cite{b3}, pathway databases such as KEGG \cite{b4}, Reactome \cite{b16}, and WikiPathways \cite{b5}, integrative collections such as MSigDB \cite{b17}, and disease-association resources such as DisGeNET \cite{b6}.

Despite its maturity, ORA still has two practical limitations. The first is limited cross-database evidence integration. Conventional ORA analyzes each database separately, leaving the comparison of their terms (database-defined gene sets such as pathways or biological processes) to the reader. Although recent tools connect or cluster terms using shared genes or semantic similarity \cite{b18,b19,b26}, their outputs often remain fragmented or coarse-grained, making it difficult to trace how related signals are connected through specific genes. As more databases are included, the cost of cross-database analysis also increases, raising scalability concerns. Second, current ORA interpretation pipelines rarely verify whether each generated biological claim is supported by the underlying enrichment evidence, a property we refer to as claim-level faithfulness. Conventional ORA returns ranked enrichment tables rather than coherent biological explanations, while visualization tools \cite{b28,b30} organize enriched terms into networks but still leave the biological reading to the user. Recent LLM- and RAG-based systems generate textual interpretations, but they are usually evaluated by semantic similarity \cite{b24}, manual inspection \cite{b23} or reproducibility \cite{b19}. This can be problematic in biomedical interpretation because unsupported gene--term associations can lead to misleading conclusions.

\begin{figure*}[t]
\centering
\includegraphics[width=0.92\textwidth]{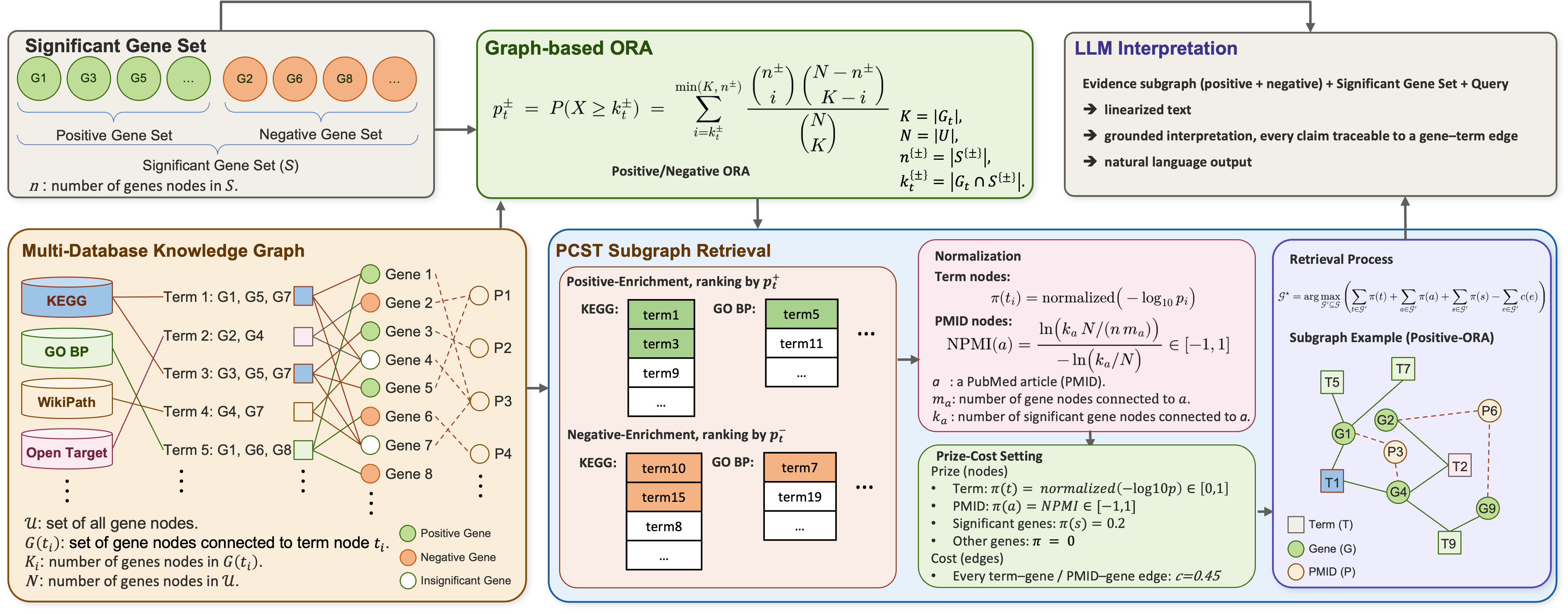}
\caption{Overview of GBOR-RAG. Significant genes are split by log-fold-change sign; each directional set is scored by graph-based ORA on the multi-database knowledge graph, enriched terms and PubMed literature nodes become node prizes, and prize-collecting retrieval extracts a compact evidence subgraph per direction that conditions the LLM interpretation.}
\label{fig:framework}
\end{figure*}

To address these two limitations, we propose \emph{Graph-Based Over-Representation RAG} (GBOR-RAG)\footnote{Our source code is available at \url{https://github.com/ggnb367/GBOR-RAG}.}, a graph-based RAG framework for ORA (Fig.~\ref{fig:framework}). For cross-database evidence integration, we formulate the task as prize-collecting subgraph extraction over a multi-database knowledge graph \cite{b11,b12}. Instead of simply concatenating the top-ranked terms from individual databases, our method retrieves a compact query-relevant subgraph that links significant terms through their shared significant genes. This representation preserves the gene-level evidence underlying each term and can surface bridge signals that connect related biological processes across databases, even when those signals are not among the top-ranked results in any single database. For quality of textual interpretation, we define a claim-level faithfulness metric to quantify whether each generated biological claim is supported by the retrieved evidence. Furthermore, we introduce an advanced variant of GBOR-RAG that adds a PubMed citation layer to the knowledge graph, providing gene--literature evidence that enriches the biological context used for textual interpretation.

We make these contributions:
\begin{itemize}
\item We implement ORA on a term--gene knowledge graph spanning 7 databases, reproducing conventional ORA with high consistency (mean Pearson $r=0.997$) while reducing query time by a median of $37\times$ (up to $48\times$).

\item We formulate cross-database ORA integration as compact connected subgraph retrieval, producing smaller evidence subgraphs (36.8 nodes on average) with full connectivity and the highest overall integration score ($4.12$).

\item We introduce a claim-level faithfulness metric for evaluating whether generated ORA interpretations are supported by their retrieved evidence, with GBOR-RAG achieving the highest score ($0.90$) among the compared systems.

\item We extend GBOR-RAG with a PubMed literature layer, improving textual interpretation quality from 4.17 to 4.75.
\end{itemize}

\section{Preliminaries}\label{sec:prelim}
\subsection{Over-Representation Analysis}

Over-representation analysis (ORA) tests whether a predefined gene set is enriched among a list of genes of interest. Let $U$ be the background gene universe with $N=|U|$, $S\subseteq U$ the significant (differentially expressed) genes with $n=|S|$, and $G\subseteq U$ the genes annotated to a term, with $K=|G|$ and $k=|G\cap S|$. ORA evaluates the one-tailed hypergeometric probability of observing at least $k$ hits by chance,
\begin{equation}
p \;=\; P(X \ge k) \;=\;
\sum_{i=k}^{\min(K,n)}
\frac{\dbinom{n}{i}\,\dbinom{N-n}{K-i}}{\dbinom{N}{K}},
\label{eq:ora-classic}
\end{equation}
and a small $p$-value marks the term as enriched; this test is the statistical core of widely used enrichment tools \cite{b1,b2}. In expression studies, significant genes are further split by the sign of their log fold change $\Delta_g$,
\begin{equation}
S^{\pm} = \{\, g \in U \mid \mathrm{adj.}p(g) < \alpha_{\mathrm{DE}},\ \pm\Delta_g > \delta \,\},
\label{eq:signed-genes}
\end{equation}
where $\delta\ge 0$ is a log fold change threshold (differential expression by limma~\cite{b7}, with $\alpha_{\mathrm{DE}}=0.05$ and $\delta=0$ in our experiments, following \cite{b8}), and ORA is run separately on $S^{+}$ and $S^{-}$ against the same $U$.

\subsection{Graph-Based Retrieval-Augmented Generation}

Retrieval-augmented generation (RAG) grounds a language model in retrieved evidence rather than parametric memory alone, reducing hallucination \cite{b9}. Graph-based RAG retrieves a subgraph whose nodes and edges the model can cite directly \cite{b10}. G-Retriever \cite{b11} formalizes this as Prize-Collecting Steiner Tree (PCST) extraction \cite{b12}, selecting a compact connected subgraph that collects high-value nodes while paying for the edges that connect them. We adopt this prize-collecting view but retrieve a general connected evidence subgraph, rather than a spanning tree, from the term--gene graph, and condition the LLM on it.

\section{Related Work}\label{sec:related}
\textbf{Cross-database integration.}
Enrichr-KG~\cite{b18} links terms across libraries and EnrichGT~\cite{b26} groups pathways with overlapping gene content, providing useful term-centered networks or clusters for exploring multi-database outputs. However, when the selected terms are determined by fixed top-$k$ rankings, the resulting evidence may remain fragmented or may not include intermediate bridge terms that connect related signals across databases. In contrast, we retrieve a compact, connected, gene-anchored subgraph that explicitly represents these cross-database links.

\textbf{Natural-language interpretation.}
Beyond visualization tools~\cite{b30,b28}, recent LLM-based systems move toward automated textual interpretation: GeneAgent~\cite{b24} self-verifies claims against curated databases, MAPA~\cite{b19} annotates clustered pathway modules, and Hu \textit{et al.}~\cite{b23} benchmark LLMs for gene-set function interpretation, but claim support is rarely evaluated against the exact evidence provided to the model. We therefore condition the LLM on the retrieved gene--term subgraph and assess claim-level faithfulness with respect to that evidence.

\section{Graph-Based Over-Representation RAG}\label{sec:method}
Those limitations of cross-database integration and quality of interpretation partly stem from the database-wise design of conventional ORA: biologically related terms across databases may share genes, but such links are not directly represented in isolated enrichment outputs. To address this, we build a multi-database term--gene knowledge graph and retrieve a compact, query-relevant subgraph that connects enriched terms through shared genes. This graph-structured evidence provides explicit support for LLM-based interpretation, improving traceability and enabling claim-level faithfulness to be assessed.

This section follows the pipeline: we build the term--gene graph (Section~\ref{sec:kg}), formulate integration as subgraph retrieval on multi-database knowledge graph (Section~\ref{sec:problem}), score terms by graph-based ORA (Section~\ref{sec:ora}), convert the scores into node prizes and retrieve a compact connected evidence subgraph by prize-collecting optimization (Section~\ref{sec:pcst}), and condition an LLM on it to generate a grounded interpretation (Section~\ref{sec:interp}). Significant genes are split by the sign of their log fold change and one subgraph is retrieved per pole, so the interpretation contrasts terms enriched among up- and down-regulated genes. We call this base pipeline \emph{naive GBOR-RAG}, and extend its retrieval with a PubMed literature layer (Section~\ref{sec:mesh}) to obtain \emph{advanced GBOR-RAG}.

\subsection{Multi-Database Knowledge Graph}\label{sec:kg}
We construct a heterogeneous multi-database knowledge graph that integrates the gene-set resources listed in Table~\ref{tab:data}, covering pathways, biological processes, disease associations, and drug-target associations. The curated libraries are taken from Enrichr ~\cite{b32}, and drug-target associations from Open Targets Platform release 25.03 ~\cite{b31}.

For each source $s\in\mathcal{D}$, let $\mathcal{T}_s$ be its terms and $A_s(t)$ the genes annotated to term $t$. We represent all annotations as a bipartite graph $\mathcal{G}=(V_T\cup V_G,E)$, where $V_T=\{(t,s)\mid t\in\mathcal{T}_s\}$ denotes the set of source-specific term nodes, $V_G$ denotes the set of gene nodes, and $E=\{\{(t,s),g\}\mid g\in A_s(t)\}$ encodes term--gene membership.

Genes are identified by Entrez ID, with HGNC symbols retained as metadata.
The source attribute $s$ distinguishes terms from different databases, while shared gene nodes provide the connections through which evidence from different sources can meet. The resulting graph contains $33{,}199$ terms, $20{,}604$ genes, and $1{,}100{,}511$ term--gene memberships.

\begin{table}[htbp]
\caption{Composition of the knowledge graph. Genes/term: mean number of gene neighbors per term or PMID; bottom rows: PubMed layer of advanced GBOR-RAG (Section~\ref{sec:mesh}).}
\label{tab:data}
\centering
\resizebox{\columnwidth}{!}{%
\begin{tabular}{llrrr}
\toprule
\textbf{Database} & \textbf{Content} & \textbf{Terms/PMIDs} & \textbf{Edges} & \textbf{Genes/term} \\
\midrule
Open Targets~\cite{b31}    & Drug-target assoc.\   & 17{,}457 &   443{,}777 &  25.4 \\
DisGeNET~\cite{b6}         & Gene-disease assoc.\  &  6{,}621 &   266{,}752 &  40.3 \\
GO BP~\cite{b3}            & Biological processes  &  6{,}036 &   197{,}436 &  32.7 \\
Reactome~\cite{b16}        & Reaction pathways     &  2{,}093 &   126{,}227 &  60.3 \\
WikiPathways~\cite{b5}     & Community pathways    &    622 &    26{,}663 &  42.9 \\
KEGG Pathway~\cite{b4}     & Molecular pathways    &    320 &    32{,}347 & 101.1 \\
MSigDB Hallmark~\cite{b17} & Hallmark gene sets    &     50 &     7{,}309 & 146.2 \\
\midrule
\textbf{Total (curated)}   &                       & 33{,}199 & 1{,}100{,}511 & 33.2 \\
\midrule
PubMed layer~\cite{b35}    & Gene citations (PMIDs) & 799{,}126 & 1{,}991{,}461 & 2.5 \\
\textbf{Combined}          &                       & 832{,}325 & 3{,}091{,}972 & 3.7 \\
\bottomrule
\end{tabular}}
\end{table}

\subsection{Problem Formulation}\label{sec:problem}

\paragraph{Problem.}
After ORA assigns each term $t$ a significance value $p_t$ with respect to the significant gene set $S$, the resulting signals remain scattered across heterogeneous databases. Because each database yields an independent ranked list, it is unclear whether terms from different sources support the same underlying mechanism. Our goal is therefore to integrate these separate enrichment signals across databases.

\paragraph{Why direct merging fails.}
Directly merging terms by raw $p$-value is unsuitable because database size, term granularity, and annotation density can bias significance distributions, causing a global ranking to favor particular resources rather than balanced cross-database evidence.

\paragraph{Formulation.}
We formulate cross-database integration as retrieving an evidence subgraph $\mathcal{G}_{\mathrm{ev}}\subseteq\mathcal{G}$ from the knowledge graph $\mathcal{G}$, containing term nodes relevant to $S$ and connecting evidence from different databases through shared genes. We require $\mathcal{G}_{\mathrm{ev}}$ to satisfy three objectives:

\begin{itemize}
    \item \textbf{Relevance.} The subgraph should prioritize terms strongly associated with $S$ while keeping heterogeneous databases balanced, rather than ranking all terms by raw $p$-values.

    \item \textbf{Connectivity.} The subgraph should connect terms through shared genes, so that cross-database integration reflects overlapping gene-level evidence rather than independent ranked lists.

    \item \textbf{Compactness.} The subgraph should avoid unnecessary expansion, keeping the retrieved evidence focused and easier to interpret and verify.
\end{itemize}
\noindent
\noindent

\subsection{Graph-Based ORA}\label{sec:ora}
To drive the retrieval, every term needs a relevance score for the significant genes, which we obtain by computing ORA directly on the knowledge graph. Each term node already carries its gene set as its neighborhood in the graph, so we reuse the hypergeometric test of Eq.~\eqref{eq:ora-classic} and only replace the predefined gene set of a term with this neighborhood
\begin{equation}
G_{t,s}=\{g\in U\mid \{(t,s),g\}\in E\},
\label{eq:neighbors}
\end{equation}
where the universe $U$ is the intersection of the genes in the KG and the measured genes. 

The significant genes are split by the sign of their log fold change into a positive set $S^{+}$ and a negative set $S^{-}$, and we score each term against both sets with the same upper-tail test: the significance $p_t^{\pm}$ of term $t$ at each pole is Eq.~\eqref{eq:ora-classic} with $G=G_{t,s}$ and $S=S^{\pm}$.

As established in the problem formulation, the $p_t^{\pm}$ are not comparable across databases, so we do not pool all terms into a single global ranking over the knowledge graph. Instead we rank the terms of each database separately, giving every term a rank within its own source. This within-database ranking is the comparable signal that Section~\ref{sec:pcst} uses to select candidate terms for retrieval. We rank by nominal $p$; Benjamini--Hochberg adjustment within a database is monotone and leaves this ranking unchanged.

\subsection{Evidence Subgraph Retrieval}\label{sec:pcst}

To assemble ranked terms into a single gene-anchored structure, we formulate retrieval as a prize-collecting subgraph problem. Term nodes receive prizes according to enrichment relevance, significant genes receive small connector prizes, and term--gene edges carry a uniform cost. The objective favors subgraphs that collect high-prize terms while retaining only the genes and edges needed to connect them, thereby removing weakly relevant or isolated nodes and producing a compact cross-database evidence core.

We instantiate this as follows. From each database we take its top-ranked terms (Section~\ref{sec:ora}) as candidate term nodes, so that every database offers its strongest evidence and none floods the graph by sheer size. Cross-database balance therefore comes from this per-database selection, and the significance-based prize below only grades candidates within the balanced pool. A candidate term $t$ receives a prize graded by its significance,
\begin{equation}
\pi(t) \;=\; \frac{\min\!\bigl(c_{\max},\, -\log_{10} p_t\bigr)}{Z},
\label{eq:termprize}
\end{equation}
where the cap $c_{\max}$ limits how much a single term can contribute and $Z$ is the normaliser defined below. A significant gene receives a small fixed prize $\pi_0$ so it can act as a connector, any other gene receives no prize, and every term-gene edge carries a uniform cost $c_0$.

We cap $-\log_{10}p_t$ at $c_{\max}$ to prevent a single extremely significant term from dominating the retrieval, and normalize by $Z=\max_t\min(c_{\max},-\log_{10}p_t)$ so that term prizes lie in $[0,1]$ and are comparable with the gene connector prize $\pi_0$.

The retrieved evidence subgraph is the connected subgraph that maximises the collected prize minus the connection cost,
\begin{equation}
\mathcal{G}_{\mathrm{ev}} \;=\; \arg\max_{\mathcal{G}'\subseteq\mathcal{G}}\ \Bigl(\sum_{v\in \mathcal{G}'}\pi(v)\;-\;\sum_{e\in \mathcal{G}'}c(e)\Bigr).
\label{eq:pcst}
\end{equation}
We solve this objective as an unrooted, single-tree PCST with the Goemans--Williamson-based \texttt{pcst\_fast} solver with strong pruning~\cite{b12,b34}, and induce the evidence subgraph from the selected nodes.

We run this retrieval once for each pole with the same prizes and costs. The positive pole uses the positive rankings with the positive genes $S^{+}$ as connectors and returns $\mathcal{G}_{\mathrm{ev}}^{+}$, and the negative pole uses the negative rankings with the negative genes $S^{-}$ and returns $\mathcal{G}_{\mathrm{ev}}^{-}$. Both subgraphs are passed to the interpretation stage of Section~\ref{sec:interp}.

\subsection{Graph Retrieval Augmented Interpretation}\label{sec:interp}
We condition a large language model on both subgraphs $\mathcal{G}_{\mathrm{ev}}^{+}$ and $\mathcal{G}_{\mathrm{ev}}^{-}$ and constrain it to ground every statement in this retrieved evidence, to promote claim-level faithfulness.

We linearize each subgraph into a textual evidence block. Within each database the terms are listed in order of their prize, and every term is followed by the significant genes that connect it, which form the overlap driving its enrichment. Because a flat listing hides the topology of the subgraph, we add two structural cues read directly off the graph, the hub genes that connect the most terms and the bridging genes that link terms from different databases. 

The positive and negative blocks are presented in parallel, so the model reports the processes enriched among the positive genes and, separately, those enriched among the negative genes, and contrasts the two directions. Both blocks are genuine enrichment signals rather than a presence-versus-absence comparison.

\subsection{Advanced GBOR-RAG}\label{sec:mesh}
Naive GBOR-RAG uses only curated gene-set libraries. To add literature context, we augment the knowledge graph with a PubMed layer from NCBI, restricted to human genes, where each article is represented as a PMID node connected to the genes it cites. Articles citing multiple significant genes can link these genes with disease, pathway, or drug-related context beyond curated terms. Table~\ref{tab:data} (bottom rows) reports the scale of this layer, which covers 26{,}309 genes (26{,}506 in the combined graph).

Unlike curated terms, PMID nodes do not define annotated gene sets and therefore cannot be scored by ORA. To bring it into the prize-collecting retrieval of Section~\ref{sec:pcst}, we need a prize for each PMID that scales with the term prizes and reflects how specifically the article relates to the significant genes.

We obtain this prize from the normalized pointwise mutual information (NPMI) between being cited by an article and being significant~\cite{b33}. With $N$ genes in the universe, $n$ of them significant, and an article citing $m$ universe genes of which $k$ are significant,
\begin{equation}
\mathrm{NPMI} \;=\; \frac{\ln\!\bigl(kN/(nm)\bigr)}{-\ln\!\bigl(k/N\bigr)} \;\in\; [-1,1].
\label{eq:npmi}
\end{equation}
The denominator normalizes the association to the fixed range $[-1,1]$, putting a PMID prize on the same bounded scale as the normalized term prizes of Section~\ref{sec:pcst}. NPMI rewards a small, focused article whose cited genes are mostly significant and penalizes a promiscuous article, whose large gene set makes the overlap expected by chance, so large screens and reviews are down-weighted automatically.

The scored PMIDs are added to the retrieval as additional prize nodes, each connected to the significant genes it co-cites, with its NPMI as the prize. The retrieval then returns a richer evidence subgraph in which the curated terms and their genes are accompanied by the primary articles that link the driver genes, providing citable mechanistic support for the interpretation beyond the curated databases.

\begin{table}[t]
\caption{Agreement between graph-based and conventional ORA, averaged over eight queries and both directions.}
\label{tab:accuracy}
\centering
\small
\setlength{\tabcolsep}{4pt}
\begin{tabular}{@{}lcccc@{}}
\toprule
\textbf{Database} & \textbf{Pearson} & \textbf{Spearman} & \textbf{Jaccard@10} & \textbf{F1} \\
\midrule
KEGG Pathway          & 0.998 & 0.996 & 0.932 & 0.926 \\
GO BP                 & 0.999 & 0.979 & 0.934 & 0.944 \\
WikiPathways          & 0.996 & 0.989 & 0.913 & 0.879 \\
Reactome              & 0.999 & 0.988 & 0.943 & 0.936 \\
MSigDB Hallmark       & 0.994 & 0.988 & 0.898 & 0.800 \\
DisGeNET              & 0.998 & 0.980 & 0.343 & 0.727 \\
Open Targets          & 0.998 & 0.981 & 0.873 & 0.709 \\
\midrule
\textbf{Overall}      & \textbf{0.997} & \textbf{0.986} & \textbf{0.834} & \textbf{0.846} \\
\bottomrule
\end{tabular}
\end{table}

\section{Experiments}\label{sec:exp}

We evaluate GBOR-RAG on ORA accuracy and efficiency, cross-database integration, claim-level faithfulness, and literature augmentation, under the following common setup. All 42 real-world queries are from GEO2KEGG~\cite{b8}, with the Alzheimer's disease query (GSE1297) used as a running example. Evidence-subgraph retrieval uses a single fixed configuration across all experiments: the top $k=20$ candidate terms per database, prize cap $c_{\max}=20$, gene connector prize $\pi_0=0.2$, and edge cost $c_0=0.45$. All interpretations were generated by the open-weight DeepSeek-V4-Pro (API model \texttt{deepseek-v4-pro}, temperature 0.2), while GPT-5-mini (\texttt{gpt-5-mini-2025-08-07}, temperature 0) served as the independent LLM judge. The LLM-as-a-judge~\cite{b21} rubrics were reviewed by bioinformatics experts and calibrated on small blinded human-rated sets; thus, the scores are used as comparative evidence under the same rubric rather than as absolute measures of biological correctness. Differences between methods are tested with two-sided paired Wilcoxon signed-rank tests over queries. Full experimental details are provided in the GitHub repository.

\subsection{Accuracy and Efficiency of Graph-based ORA}\label{sec:rq1}

\begin{figure}[t]
\centering
\includegraphics[width=\columnwidth]{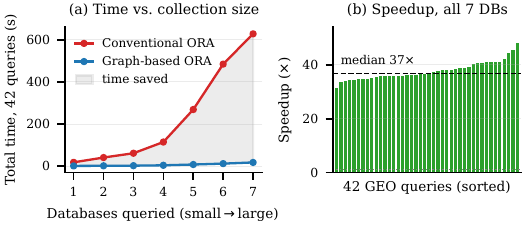}
\caption{Query-time efficiency on the 42 GEO queries. (a) Total query time as databases are added from smallest to largest, with shaded savings from Graph-based ORA. (b) Per-query speedup on all seven databases (sorted), with the dashed line marking the median.}
\label{fig:rq1eff}
\end{figure}

\textbf{Purpose.}
This experiment checks that graph-based ORA is a sound foundation: scoring ORA on the graph should not change the result (accuracy) and scale to many databases (efficiency).

\textbf{Settings.}
Efficiency uses all 42 queries of GEO2KEGG and the seven databases of Table~\ref{tab:data}; accuracy compares against conventional ORA from GSEABenchmarkeR~\cite{b8} on the eight queries with a reference, both sharing the same significant genes and size filter. Accuracy is measured over query-database-direction pairs by Pearson and Spearman correlation of $-\log_{10}(p)$, by Jaccard@10 on the top-10 terms, and by F1 between the sets of terms with nominal $p<0.05$; efficiency is per-query time after the one-time graph build, adding databases cumulatively.

\textbf{Accuracy and efficiency.}
Graph-based ORA closely reproduces conventional ORA (Table~\ref{tab:accuracy}), with an overall Pearson correlation of $0.997$ and database-level correlations of $0.994$--$0.999$; lower top-term overlap in DisGeNET and Open Targets mainly reflects differences in library version and background universe between the two implementations rather than in the test itself. Because it uses the same hypergeometric test on a pre-built graph index, it preserves conventional ORA while improving efficiency: on all seven databases, query time drops from about $15.6$ s to $0.4$ s, a median $37\times$ speedup (up to $48\times$) that grows as databases are pooled (Fig.~\ref{fig:rq1eff}).

\subsection{Quality of Cross-Database Integration}\label{sec:integration}

\textbf{Purpose.}
We formulate cross-database integration result as retrieving an evidence subgraph that links terms from multi-databases through shared genes, and we ask whether the retrieved subgraph is \emph{relevant}, cross-database \emph{connected}, and \emph{compact} (Section~\ref{sec:problem}). 

\textbf{Settings.}
We evaluate GBOR-RAG retrieval subgraph on 42 queries against two baselines derived from the same ORA scores and per-database candidate terms. Top-$k$ pooling aggregates the highest-ranked terms from each database without modeling cross-database connections, whereas Enrichr-KG \cite{b18} links enriched terms through genes shared across libraries. We evaluate each retrieved evidence subgraph using structural metrics for compactness and connectivity, together with a rubric covering relevance, term-level integration, gene-level bridging, and overall usefulness as interpretation evidence.

\textbf{Judge reliability.}
To calibrate the LLM-as-a-judge evaluation, we randomly selected 3 queries and evaluated all 18 retrieved positive and negative subgraphs. Each subgraph was independently scored by two trained human annotators blinded to method identity, as well as by the LLM judge. The human annotators showed strong rank agreement (pooled Spearman $\rho=0.74$), suggesting that the rubric supports consistent human assessment. The LLM judge showed moderate agreement with the human average ($\rho=0.53$), with stronger agreement on integration and overall usefulness (Table~\ref{tab:integration-cal}). We therefore use the LLM judge for scalable aggregate comparisons, not to replace human annotation of individual subgraphs.

\textbf{Result.}
GBOR-RAG best balances compactness and integration (Table~\ref{tab:integration}); paired tests confirm its advantage over top-$k$ pooling in integration, bridging, and overall usefulness ($p<10^{-3}$), while its overall usefulness does not differ significantly from Enrichr-KG ($p=0.08$).
\begin{itemize}
    \item \textbf{Relevance.}
    Under strong compression, GBOR-RAG matches top-$k$ pooling in relevance but trails Enrichr-KG.
    \item \textbf{Connectivity.}
    Top-$k$ pooling yields fragmented subgraphs, whereas GBOR-RAG forms a single connected component for every pole and obtains the highest mean integration score.
    \item \textbf{Compactness.}
    GBOR-RAG reduces the mean subgraph size from 1466.7 (top-$k$ pooling) and 470.9 (Enrichr-KG) to 36.8 nodes.
\end{itemize}

\begin{table}[t]
\caption{Judge calibration (Spearman $\rho$) among human annotators and between their average and the LLM judge, on 18 subgraphs (top; Section~\ref{sec:integration}) and 12 interpretations (bottom; Section~\ref{sec:exp-adv}).}
\label{tab:integration-cal}
\centering
\footnotesize
\setlength{\tabcolsep}{3.5pt}
\begin{tabular}{@{}lccccccc@{}}
\toprule
\textbf{Subgraph} & \textbf{Relev.} & \textbf{Integr.} & \textbf{Bridg.} & \textbf{Overall} & & & \textbf{All} \\
\midrule
Inter-annotator & 0.59 & 0.68 & 0.85 & 0.68 & & & \textbf{0.74} \\
Human--LLM      & 0.51 & 0.61 & 0.59 & 0.64 & & & \textbf{0.53} \\
\midrule
\textbf{Interpretation} & \textbf{Rel.} & \textbf{Mec.} & \textbf{Int.} & \textbf{Anc.} & \textbf{Cla.} & \textbf{Ovr.} & \textbf{All} \\
\midrule
Inter-annotator & 0.62 & 0.75 & 0.62 & 0.53 & 0.56 & 0.45 & \textbf{0.67} \\
Human--LLM      & 0.65 & 0.59 & 0.74 & 0.62 & 0.77 & 0.74 & \textbf{0.69} \\
\bottomrule
\end{tabular}
\end{table}

\begin{table}[t]
\caption{Cross-database integration (mean per pole). Structure: nodes, connected components (Comp.), and fraction of single-component poles (single\_cc). Content: LLM-judge rating ($1$--$5$).}
\label{tab:integration}
\centering
\footnotesize
\setlength{\tabcolsep}{3pt}
\begin{tabular}{@{}lrrrcccc@{}}
\toprule
 & \multicolumn{3}{c}{\textbf{Structure}} & \multicolumn{4}{c}{\textbf{Content}} \\
\cmidrule(lr){2-4}\cmidrule(l){5-8}
\textbf{Method} & \textbf{Nodes} & \textbf{Comp.} & \textbf{single\_cc} & \textbf{Rel.} & \textbf{Int.} & \textbf{Bri.} & \textbf{Ovr.} \\
\midrule
Top-$k$ pooling   & 1466.7 & 15.64 & 0.00 & 3.89 & 3.68 & 2.52 & 3.20 \\
Enrichr-KG        & 470.9  & 1.05  & 0.95 & \textbf{4.09} & 4.29 & \textbf{4.62} & 3.92 \\
\textbf{GBOR-RAG} & \textbf{36.8} & \textbf{1.00} & \textbf{1.00} & 3.89 & \textbf{4.31} & 4.43 & \textbf{4.12} \\
\bottomrule
\end{tabular}
\end{table}

\subsection{Claim-Level Faithfulness of Textual Interpretation}\label{sec:rq2}

\textbf{Metric.} We introduce \emph{claim-level faithfulness} to quantify whether a generated interpretation is supported by its evidence. A report $R$ is decomposed into a set of atomic claims $\mathcal{C}(R)$ ~\cite{b20}, each asserting a gene, a term, a gene--term relation, or a biological interpretation. Given the evidence $E$ that the method supplied to its generator, an independent LLM judge labels each claim $c$ as supported ($s(c,E)=1$) when it is entailed by $E$ and unsupported ($0$) otherwise. Claim-level faithfulness is the fraction of supported claims,
\begin{equation}
\mathrm{Faithfulness}(R,E)=\frac{1}{|\mathcal{C}(R)|}\sum_{c\in\mathcal{C}(R)} s(c,E),
\end{equation}
averaged over all queries. The metric measures consistency with the evidence a system was given, not biological correctness.

\textbf{Setting.} On 42 queries we compare GBOR-RAG against two published gene-set interpretation systems: GeneAgent \cite{b24}, which verifies claims through database tool calls, and MAPA \cite{b19}, which clusters enriched pathways into functional modules before generation. Each system produces one interpretation from its own evidence, decomposed and judged by an independent LLM judge from a different model family than the generators.

\textbf{Result.}
As shown in Table~\ref{tab:claimfaith}, GBOR-RAG attains the highest claim-level faithfulness ($0.90$) with compact evidence, using about $27\times$ less evidence than GeneAgent and only slightly more than MAPA. GeneAgent reaches $0.71$ despite its much larger evidence context, whereas MAPA reaches $0.29$ on the subset of queries where its module-clustering step applies. GBOR-RAG also leaves the fewest unsupported claims, and its faithfulness advantage is significant over shared queries (vs.\ GeneAgent $p<10^{-5}$; vs.\ MAPA $p<10^{-6}$). These results suggest that a compact, traceable evidence subgraph improves faithfulness while avoiding the large evidence context required by agentic verification.

\begin{table}[t]
\caption{Claim-level faithfulness. Claims and Supported: mean counts per report; Faithfulness: mean per-report supported-claim ratio; Evidence: tokens.}
\label{tab:claimfaith}
\centering
\small
\setlength{\tabcolsep}{5pt}
\begin{tabular}{@{}lrrrr@{}}
\toprule
\textbf{Method} & \textbf{Evidence} & \textbf{Claims} & \textbf{Supported} & \textbf{Faithfulness} \\
\midrule
GeneAgent         & 32.33K & 19.90 & 13.88 & 0.71 \\
MAPA              & 0.80K  & 20.52 & 5.80  & 0.29 \\
\textbf{GBOR-RAG} & \textbf{1.19K} & 9.39 & 8.15 & \textbf{0.90} \\
\bottomrule
\end{tabular}
\end{table}

\begin{figure}[t]
\centering
\includegraphics[width=\columnwidth]{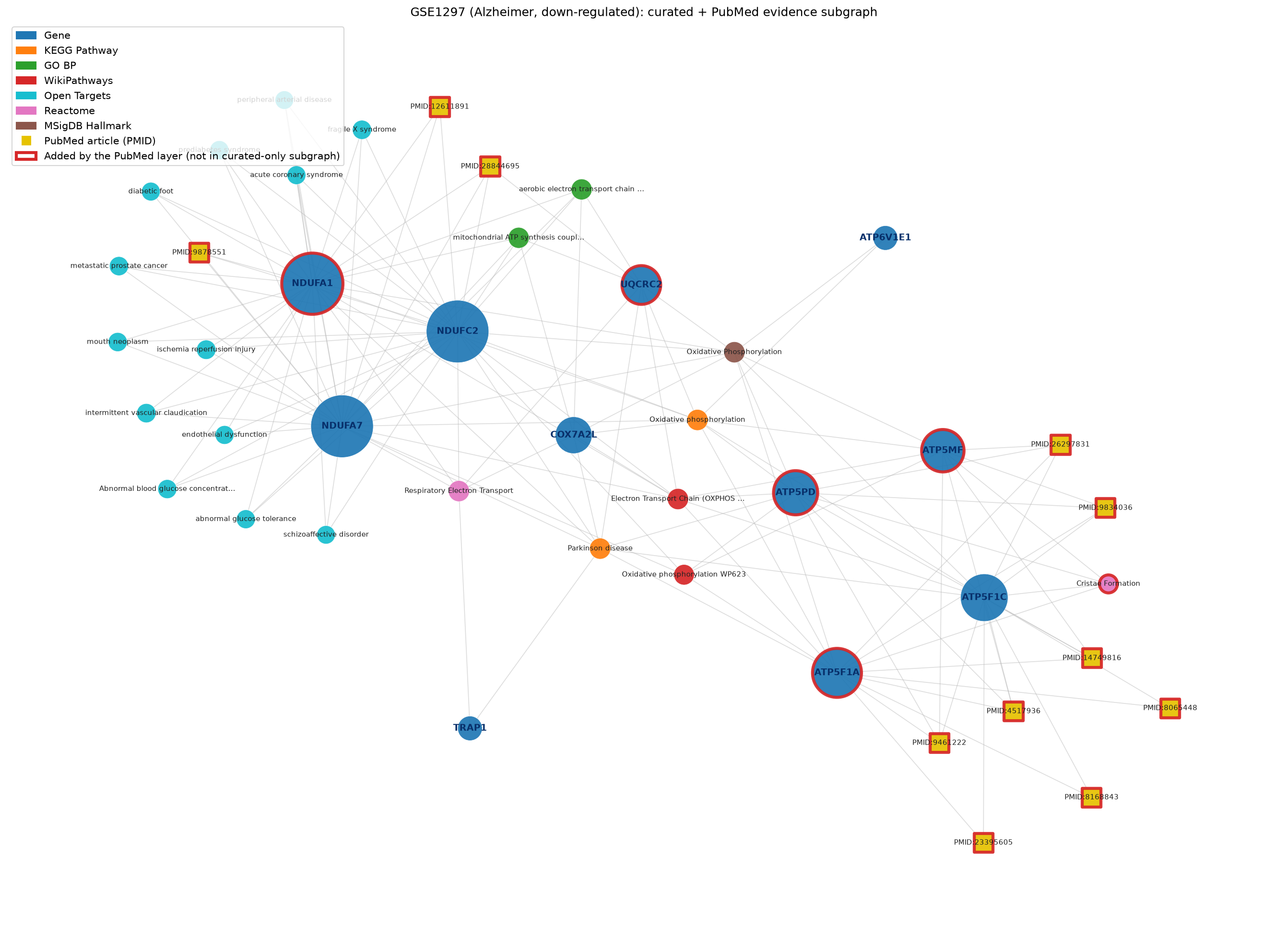}
\caption{Advanced evidence subgraph for the negative ORA of the Alzheimer's disease query. Curated terms and genes define a mitochondrial oxidative-phosphorylation core; gold squares denote PubMed articles, and crimson rings mark literature-layer nodes absent from the curated-only subgraph.}
\label{fig:advsub}
\end{figure}

\subsection{Advanced GBOR-RAG: Literature-Augmented Retrieval}
\label{sec:exp-adv}

\textbf{Purpose.}
We evaluate whether adding a PubMed literature layer improves the generated interpretation. 

\textbf{Setting.}
Naive and advanced GBOR-RAG use the same generator and prose prompt, differing only in the retrieved evidence. On 42 queries, a blind independent LLM judge rates each interpretation on six dimensions: relevance and specificity, mechanistic insight, integration and synthesis, evidence anchoring, clarity, and overall quality.

\textbf{Judge reliability.}
Three trained annotators rated a blind calibration set of 6 queries, comprising 12 interpretations, using the same rubric. Agreement was substantial among annotators (pooled Spearman $\rho=0.67$) and between the human average and the LLM judge (pooled $\rho=0.69$, overall-quality $\rho=0.74$; Table~\ref{tab:integration-cal}).

\textbf{Result.}
As shown in Table~\ref{tab:interpqual}, advanced GBOR-RAG improves all content-related dimensions while keeping clarity essentially unchanged. The overall score increases from $4.17$ to $4.75$ ($p<10^{-6}$), with the largest gains in evidence anchoring ($+0.75$) and mechanistic insight ($+0.39$). This suggests that the PubMed layer provides stronger grounding and richer mechanistic context without reducing readability.

\begin{table}[t]
\caption{Interpretation quality of naive vs.\ advanced GBOR-RAG (mean LLM-judge rating, $1$--$5$, three judgments per text; 42 queries).}
\label{tab:interpqual}
\centering
\small
\begin{tabular}{@{}lcc@{}}
\toprule
\textbf{Dimension} & \textbf{Naive} & \textbf{Advanced} \\
\midrule
Relevance \& specificity & 3.81 & \textbf{4.16} \\
Mechanistic insight      & 3.72 & \textbf{4.11} \\
Integration \& synthesis & 4.48 & \textbf{4.80} \\
Evidence anchoring       & 4.23 & \textbf{4.98} \\
Clarity                  & 4.98 & \textbf{5.00} \\
\textbf{Overall}         & 4.17 & \textbf{4.75} \\
\bottomrule
\end{tabular}
\end{table}

\textbf{Qualitative example.}
For GSE1297, as shown in Fig.~\ref{fig:advsub} for the negative subgraph, advanced GBOR-RAG incorporates PMIDs into prize-collecting retrieval, allowing it to collect richer biological evidence than naive GBOR-RAG.

\section{Conclusion}
We presented GBOR-RAG, which integrates multi-database ORA through prize-collecting retrieval on a term--gene knowledge graph and conditions an LLM on the retrieved gene-anchored evidence. Graph-based ORA reproduces conventional ORA with a median $37\times$ speedup, the retrieved subgraphs are compact and connected, the resulting interpretations are more faithful to their evidence than those of existing systems, and a PubMed literature layer further improves interpretation quality. Our evaluation relies on an LLM judge calibrated on small human-rated sets, and faithfulness reflects consistency with the evidence rather than biological correctness; a larger expert study is left for future work.

\section*{Acknowledgment}
This work was supported by JST BOOST, Japan Grant Number JPMJBS2402,
and JSPS KAKENHI Grant Number JP26K23914.

\end{document}